\documentclass[11pt]{article}
\usepackage{moriond,epsfig}

\def\be{\begin{equation}}
\def\ee{\end{equation}}
\def\bea{\begin{eqnarray}}
\def\eea{\end{eqnarray}}

\newcommand{\ks}{K^{0}_{S}}

\newcommand{\ksp}{K^{0}_{S}\> p}
\newcommand{\kspb}{K^{0}_{S}\> \bar{p}}
\newcommand{\ksppb}{K^{0}_{S}\> p\>(\bar{p})}
\newcommand{\coll}{Collaboration}
\newcommand{\etal}{ {\it et al.,} }
\newcommand{\gev}{\; \mathrm{GeV}}
\newcommand{\mev}{\; \mathrm{MeV}}

\begin{document}
%\vspace*{4cm}
\vspace*{3cm}
\title{RECENT SPECTROSCOPY RESULTS FROM ZEUS}

\author{ A.~GEISER \\ for the ZEUS Collaboration}

\address{DESY, Notkestrasse 85, D-22607 Hamburg, Germany; 
E-mail: Achim.Geiser@desy.de}

\maketitle\abstracts{Recent results on light hadron spectroscopy are reported,
with special emphasis on the evidence for a narrow baryonic state decaying
to $\ksp$ and $\kspb$, compatible with the pentaquark state $\theta^+$ 
observed by fixed target experiments.
The data were collected with the ZEUS detector at HERA using an integrated 
luminosity of 121 pb$^{-1}$. The analyses were performed in the
central rapidity region of inclusive deep inelastic scattering at
an $ep$ centre-of-mass energy of 300--318 GeV.
Evidence for a narrow resonance in the $\ksp$ and $\kspb$ invariant mass 
spectrum is obtained, with mass $1521.5 \pm 1.5(stat)^{+2.8}_{-1.7}(syst)$
and width consistent with the experimental resolution. If the $\ksp$ part
of the signal is identified with the strange pentaquark $\theta^+$, the 
$\kspb$ part is the first evidence for its antiparticle, $\bar \theta^-$.
Supporting results on other light hadron resonances are also discussed.}

\section{Introduction}
\label{intro}

Recently, interest in light hadron spectroscopy has been considerably revived
by the observation of baryonic resonances \cite{fixed,ks,NA49} 
compatible with their interpretation in terms of pentaquark states, 
i.e. bound states of four quarks and an antiquark \cite{zp:a359:305,pentath}. 
While almost all hadronic states
observed previously can be interpreted in terms of baryons (bound states of
three quarks, qqq) or mesons (bound quark-antiquark states, q\=q), the theory 
of Quantum Chromo Dynamics (QCD) does not preclude the existence of other 
colour neutral quark combinations such as tetraquarks (qq\=q\=q), pentaquarks
(qqqq\=q), etc., or states including gluons such as glueballs (gg) or
hybrids (q\=qg).

Potential glueball candidates have already been discussed for many years 
\cite{glueball} without firm conclusions. The ZEUS experiment at the HERA 
ep collider has recently added input to this discussion through the 
observation of at least one of these candidates in the $\ks$$\ks$ final state 
\cite{zeusksks}.     

The first quantitative predictions for the existence of bound pentaquark 
states have been made only fairly recently \cite{zp:a359:305}, and their 
observation, mainly of the $\theta^+$ state with quark content uudd\=s,  has 
triggered a large number of further theoretical predictions \cite{pentath}.
However, the first pentaquark observations were all obtained in 
fixed target experiments at low centre-of-mass energies \cite{fixed,ks}, where
valence quarks from the target nucleons can easily become part of the 
pentaquark final state. Consequently, only qqqq\=q states were found,
in contrast to \=q\=q\=q\=qq states.

Here we investigate the production of light narrow resonances in  
the central rapidity region of inclusive deep inelastic scattering (DIS) at
an $ep$ centre-of-mass energy of 300--318 GeV for exchanged photon
virtuality, $Q^2$, above 1 $\gev^2$.
In this kinematic region, hadrons are mainly produced from the fragmentation
of a sea (anti)quark in the proton struck by the photon, and only faintly
``remember'', if at all, their proton parent. Hadrons and their antiparticles  
 are hence produced in approximately equal proportions.
      
It is therefore a nontrivial question whether the pentaquarks observed in 
fixed target collisions should also be seen in high energy collider
experiments. On the other hand, if at all, collider 
experiments should observe not only the pentaquarks, but also their 
antiparticles. 
The potentially first such observation in the light quark sector
is presented in this document.
Corresponding experimental evidence and/or searches for pentaquarks 
containing a heavier charmed quark are discussed in \cite{pentaH1,zeusH1}. 

%%%%%%%%%%%%%%%%%%%%%%%%%%%%%%%%%%%%%%%%%%%%%%%%%%%%%%%%%%%%%
\section{Evidence for a narrow baryonic state decaying to $\ksppb$}
%%%%%%%%%%%%%%%%%%%%%%%%%%%%%%%%%%%%%%%%%%%%%%%%%%%%%%%%%%%%%

ZEUS has performed a search for pentaquarks in the $\ksppb$  decay
channel \cite{zeuspenta}. The analysis used deep inelastic scattering events
measured with exchanged-photon virtuality $Q^2\ge 1\gev^2$. The
data sample corresponds to an integrated luminosity of 121
pb$^{-1}$. The charged tracks were selected in the central
tracking detector (CTD) with $p_T\ge 0.15\gev$ and $|\eta|\le 1.75$,
restricting this study to a region where the CTD track acceptance
and resolution are high. 

\begin{center}
\begin{minipage}[c]{0.54\textwidth}
\includegraphics[width=8.5cm,angle=0]{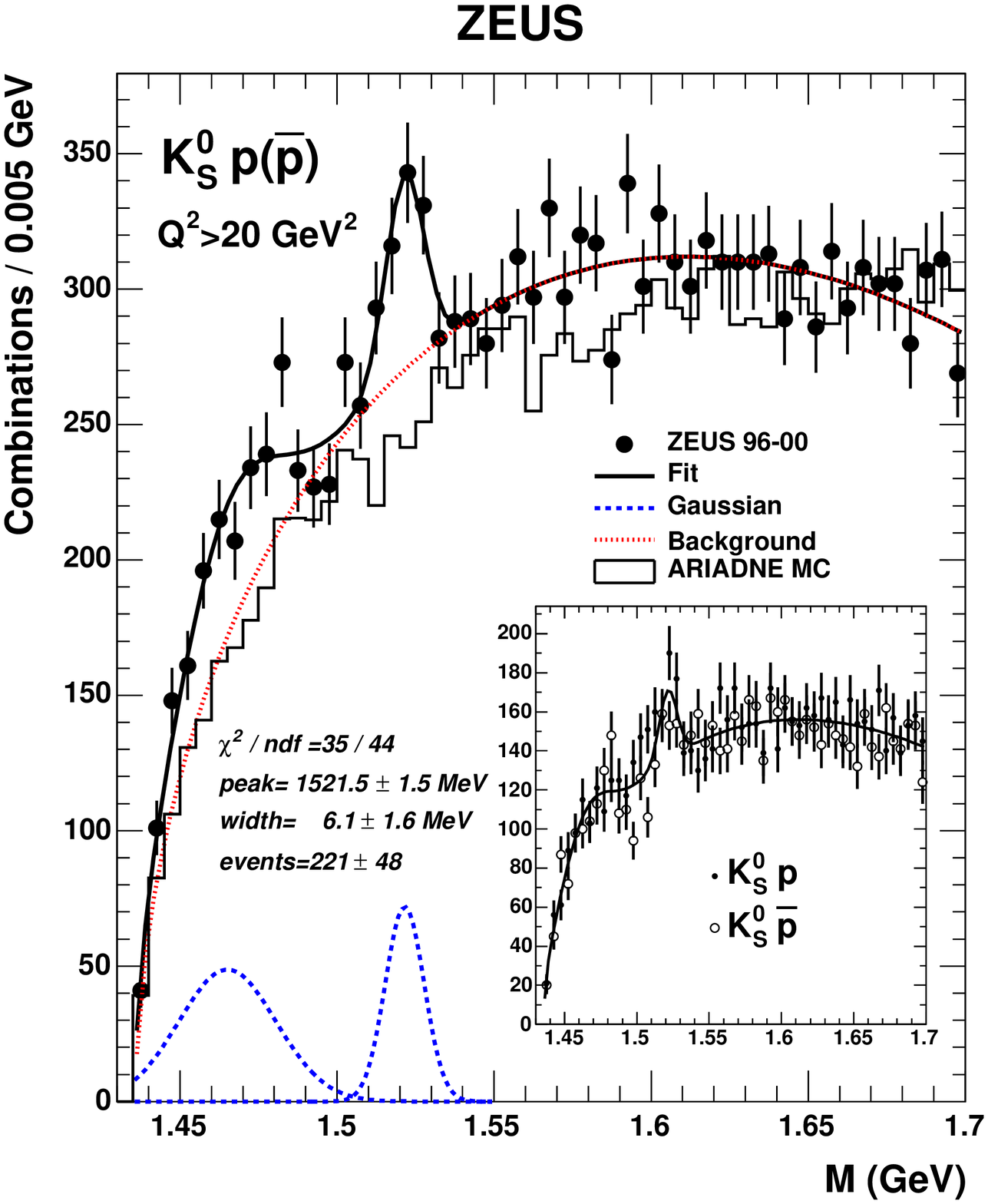}
\label{theta}
\end{minipage}
\hfill
\begin{minipage}[c]{.42\textwidth}
Figure 1: { Invariant-mass spectrum for the $\ksppb$ channel for
$Q^2 > 20 \gev^2$. The solid line is the result of a fit to the
data using a threshold background plus two Gaussians.  The
dashed lines show the Gaussian components, while the dotted line
indicates background. The prediction of the Monte Carlo
simulation is normalised to the data in the mass region above
$1650\mev$. The inset shows the $\kspb$ (open circles) and the
$\ksp$ (black dots) candidates separately, compared to the result
of the fit to the combined sample scaled by a factor of 0.5. }
\end{minipage}
\end{center}

The total number of $\ks$ with $p_{T}(\ks)>0.3\gev$ and $|\eta(\ks)|<1.5$,
identified using the decay mode $\ks\to\pi^{+}\pi^{-}$, 
was  867K. 
To eliminate contamination from $\Lambda (\bar{\Lambda})$ decays,
candidates with a proton mass hypothesis $M(p\> \pi)<1121\mev$
were rejected.

The (anti)proton-candidate selection used the energy-loss
measurement in the CTD, $dE/dx$. $\ksppb$ invariant masses were
obtained by combining $\ks$ candidates in the mass region
$480-510\mev$ with (anti)proton candidates in the (anti)proton
$dE/dx$ band with the additional requirements $p<1.5\gev$ and
$dE/dx>1.15$ mips in order to reduce the pion background.  The CTD
resolution for the $\ksppb$ invariant-mass near $1530\mev$,
was estimated from Monte Carlo simulations to be 
$2.0\pm 0.5\mev$ for
both the $\ksp$ and the $\kspb$ channels.

Fig.~1  shows the $\ksppb$ invariant mass  for $Q^2> 20\gev^2$.
The separated $\ksp$ and $\kspb$ distributions are shown as
inset. 
The choice of the higher $Q^2$ cut was motivated by the expected decrease
in acceptance and increase in background at lower $Q^2$ values.
Mass distributions for different $Q^2$ cuts can be found in \cite{zeuspenta}.
The fit of two Gaussians on top of a continous background, also shown in 
Fig. 1, yields a clear narrow peak at $1522\mev$. The second Gaussian
near $1470\mev$ could either be attributed to another, broader resonance 
(e.g. the unestablished resonance $\Sigma (1480)$ \cite{PDG}), or 
be regarded as the empirical parametrization of a more complicated
background shape. A single Gaussian fit yields essentially
identical results \cite{zeuspenta}, indicating robustness against the 
choice of the background function.   
From the {\sc Ariadne} Monte Carlo model, which contains only ordinary
hadrons, no peaks are expected in this region. 

The peak position obtained from Fig. 1 is $1521.5\pm 1.5({\rm
stat.})^{+2.8}_{-1.7} ({\rm syst.}) \mev$ (see \cite{zeuspenta} for details). 
It agrees well with the
measurements by HERMES, SVD and COSY-TOF for the same decay
channel \cite{ks}, but is slightly lower than the mass found in experiments
reconstructing the $K^+ n$ decay channel \cite{fixed}. 
Its Gaussian width was found to be $6.1 \pm 1.6$~GeV,
still compatible with the experimental resolution of 2 MeV.
If the width of the Gaussian is fixed to this
experimental resolution, the extracted Breit-Wigner width of the
signal is $\Gamma=8\pm 4(\mathrm{stat.}) \mev$.

The number of signal events obtained from the fit in Fig. 1 is $221
\pm 48$.  This corresponds to a statistical significance of $4.6\sigma$. The
number of events in the $\kspb$ channel is $96 \pm 34$. It agrees
with the signal extracted for the $\ksp$ decay mode.
If the $\ksp$ signal corresponds to the $\theta^+$ pentaquark observed
by other experiments, this measurement 
provides the first evidence for its antiparticle with a quark
content of $\bar{u}\bar{u}\bar{d}\bar{d}s$.

\section{Other resonances and systematic checks}

Searches for corresponding double strange and charmed pentaquark states
have also been performed by ZEUS. The corresponding (negative) results
can be found in \cite{zeusss} and \cite{zeusH1}, contrasting the positive 
evidence 
obtained by the NA49 \cite{NA49} and H1 \cite{pentaH1} collaborations. 

\begin{center}
\begin{minipage}[c]{0.58\textwidth}
\includegraphics[width=9.5cm,angle=0]{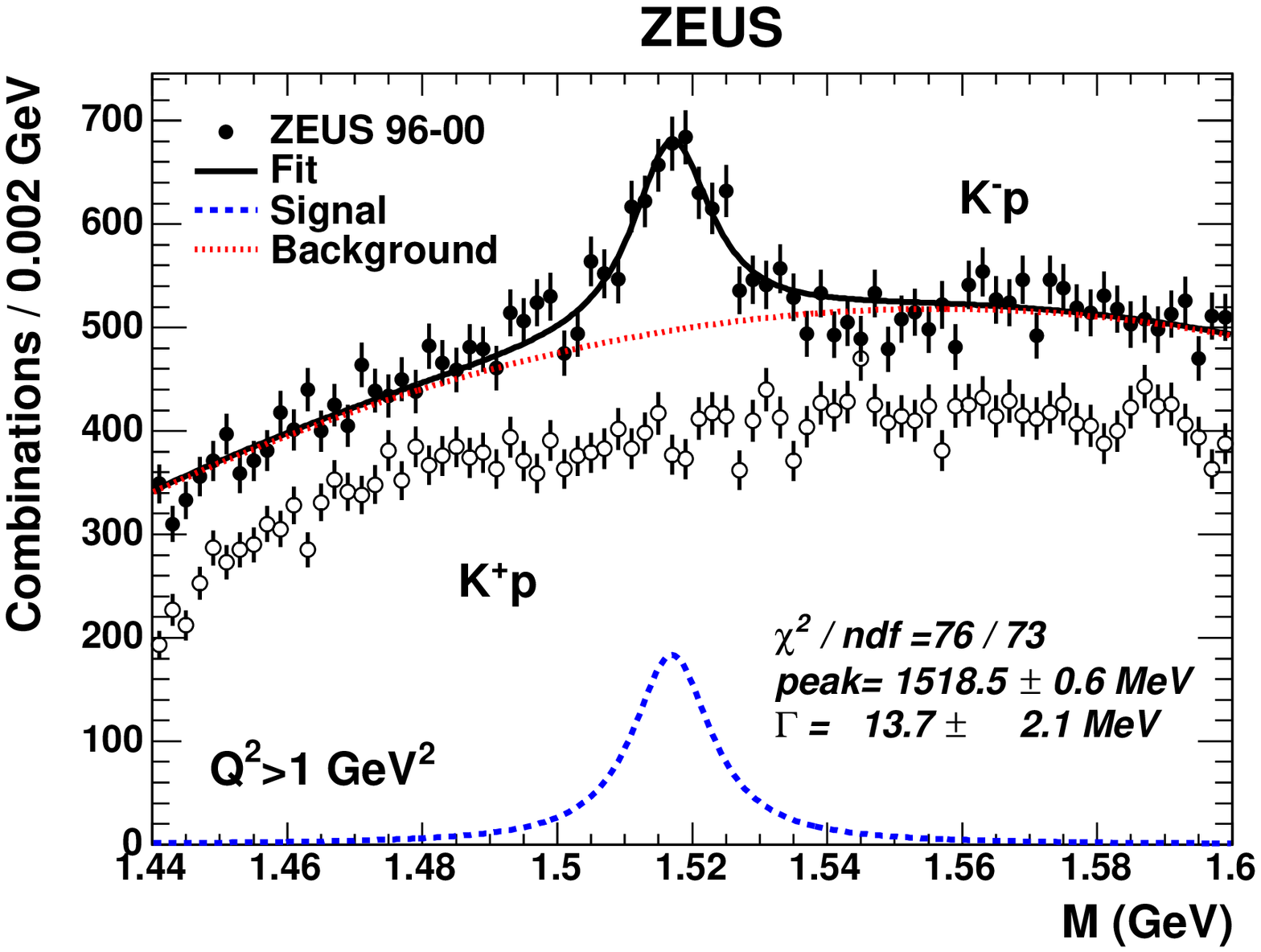}
\label{theta++}
\end{minipage}
%
% \hfill
\begin{minipage}[c]{.40\textwidth}
Figure 2: { Invariant-mass spectra for the $K^+p$ and $K^-p$
channels (plus charge conjugates) for $Q^2 > 1 \gev^2$. }
\end{minipage}
\end{center}

Previous studies on $K^0$ and $\Lambda$ production in DIS \cite{zeusksks} 
enhance the 
confidence in the $K^0$ reconstruction and selection procedure, and allow 
cross-checks of proton identification.
Events with two $K^0$'s have been successfully used to study resonances
decaying into $K^0_s$ pairs \cite{zeusksks}.
Many further systematic checks are explained in \cite{zeuspenta}.

For example, an important contribution to understand and check the 
efficiencies and 
resolutions can be obtained from the reconstruction of the well established 
$\Lambda(1520)D_{03}$ meson in its charged $pK$ decay mode. 
The $K^\pm
p (K^\pm\bar{p})$ invariant mass spectra were investigated in  a
wide range of minimum $Q^2$ values, identifying proton and charged
kaon candidates using the $dE/dx$ information. The proton
candidates inside the $dE/dx$ proton band were required to have
$dE/dx>1.8$ mips, while the kaon candidates were reconstructed in
the kaon band after the restriction $dE/dx >1.2$ mips. For
$Q^2>1\gev^2$, no peak was observed near 1522 MeV in the $K^+p$
and $K^-\bar{p}$ spectra, see Fig.~2, while a clean signal was seen in
the $K^-p (K^+\bar{p})$ channel at $1518.5\pm
0.6(\mathrm{stat}.)\mev$, corresponding to the
$\Lambda(1520)D_{03}$ state. The fact that both mass and width are
in good agreement with the PDG\cite{PDG} values verifies that there 
are no significant reconstruction biases in the pentaquark mass range.
Also, the absence of a peak in the like sign spectrum disfavours the isotensor
interpretation of the $\theta^+$ state \cite{plb570:185}.

\section{Conclusions}
Evidence for a narrow baryonic state decaying
to $\ksp$ and $\kspb$ is obtained in the central fragmentation region of 
deep inelastic scattering events by the ZEUS collaboration at HERA.
This state is compatible with the pentaquark state $\theta^+$ 
observed by fixed target experiments, and its antiparticle $\bar\theta^-$,
presumably observed for the first time. 
Supporting results on other light hadron resonances are also presented.

%\section*{Acknowledgments}
%This is where one places acknowledgments for funding bodies etc.
%Note that there are no section numbers for the Acknowledgments, Appendix
%or References.

\end{document}